%% file: paper.tex
\documentclass[sigconf, anonymous=False, dvipsnames, table]{acmart}
\AtBeginDocument{%
  }

\usepackage{bm}
\usepackage{enumitem}
\usepackage{graphicx}
\usepackage{pifont}
\usepackage{textcomp}
\usepackage{xcolor}
\usepackage{booktabs}
\usepackage{todonotes}
\usepackage{listings}
\usepackage{minted}
\usepackage{annotate-equations}
\usepackage{cleveref}
\usepackage{subcaption}
\usepackage{tablefootnote}
\usepackage{xspace}

\newcommand{\rev}[1]{#1}

\newcommand{\gchk}{\cellcolor{green!25}\checkmark}
\newcommand{\yque}{\cellcolor{yellow!25}?}
\newcommand{\bcq}{\cellcolor{brown!25}\checkmark?}

\newcommand{\benchsuite}{\textsc{ParEval-Repo}\xspace}

\definecolor{codegreen}{rgb}{0,0.6,0}
\definecolor{codegray}{rgb}{0.5,0.5,0.5}
\definecolor{codepurple}{rgb}{0.58,0,0.82}
\definecolor{backcolour}{rgb}{0.95,0.95,0.92}

\lstdefinestyle{mystyle}{
    backgroundcolor=\color{backcolour},
    commentstyle=\color{codegreen},
    keywordstyle=\color{magenta},
    numberstyle=\tiny\color{codegray},
    stringstyle=\color{codepurple},
    basicstyle=\ttfamily\footnotesize,
    breakatwhitespace=false,
    breaklines=true,
    captionpos=b,
    keepspaces=true,
    numbers=left,
    numbersep=5pt,
    showspaces=false,
    showstringspaces=false,
    showtabs=false,
    tabsize=2
}
\definecolor{myblue}{RGB}{63, 90, 126}
\definecolor{mygray}{RGB}{228, 244, 247}

\usepackage{xparse}
\NewDocumentCommand{\rot}{O{45} O{1em} m}{\makebox[#2][l]{\rotatebox{#1}{#3}}}

\copyrightyear{2025}
\acmYear{2025}
\setcopyright{cc}
\setcctype{by}
\acmConference[ICPP '25]{54th International Conference on Parallel Processing}{September 08--11, 2025}{San Diego, CA, USA}
\acmBooktitle{54th International Conference on Parallel Processing (ICPP '25), September 08--11, 2025, San Diego, CA, USA}
\acmPrice{}
\acmDOI{10.1145/3754598.3754669}
\acmISBN{979-8-4007-2074-1/25/09}

\begin{document}

\title{\benchsuite: A Benchmark Suite for Evaluating LLMs with Repository-level HPC Translation Tasks}

\author{Joshua H. Davis}
\affiliation{%
  \institution{Department of Computer Science,\\University of Maryland}
  \city{College Park}
  \state{Maryland}
  \country{USA}
}
\email{jhdavis@umd.edu}

\author{Daniel Nichols}
\affiliation{%
  \institution{Department of Computer Science,\\University of Maryland}
  \city{College Park}
  \state{Maryland}
  \country{USA}
}
\email{dnicho@umd.edu}

\author{Ishan Khillan}
\affiliation{%
  \institution{Department of Computer Science,\\University of Maryland}
  \city{College Park}
  \state{Maryland}
  \country{USA}
}
\email{ikhillan@umd.edu}

\author{Abhinav Bhatele}
\affiliation{%
  \institution{Department of Computer Science,\\University of Maryland}
  \city{College Park}
  \state{Maryland}
  \country{USA}
}
\email{bhatele@cs.umd.edu}

\renewcommand{\shortauthors}{Davis et al.}

\begin{abstract}
\input{abstract}
\end{abstract}


\maketitle

\section{Introduction}
\label{sec:intro}
\input{intro}

\section{Background}
\label{sec:bg}
\input{bg}

\section{Techniques for Repo-level Translation}
\label{sec:techniques}
\input{techniques}

\section{Large Language Models Evaluated}
\label{sec:llms}
\input{llms}

\section{The \benchsuite{} Suite of Translation Tasks}
\label{sec:benchmarks}
\input{benchmarks}

\section{Metrics for Repo-level Translation Correctness and Performance}
\label{sec:metrics}
\input{metrics}

\section{Experimental Setup}
\label{sec:setup}
\input{setup}

\section{Results}
\label{sec:results}
\input{results}

\section{Related Work}
\label{sec:related}
\input{related}

\section{Conclusion}
\label{sec:conc}
\input{conc}

\begin{acks}
\input{ack}
\end{acks}

\bibliographystyle{ACM-Reference-Format}
\bibliography{./bib/pssg,./bib/cite}

\end{document}

%% file: abstract.tex
GPGPU architectures have become significantly \rev{more} diverse in recent
years, which has led to an emergence of a variety of specialized programming
models and software stacks to support them. Portable programming models exist,
but they require significant developer effort to port to and optimize
for different hardware architectures. Large language models
(LLMs) may help to reduce this programmer burden. In this paper, we present
a novel benchmark and testing framework, \benchsuite, which can be used to
evaluate the efficacy of LLM-based approaches in automatically translating
entire codebases across GPGPU execution models. \benchsuite includes several
scientific computing and AI mini-applications in a range of programming models
and levels of repository complexity. We use \benchsuite to evaluate a range of
state-of-the-art open-source and \rev{commercial} LLMs, with both a non-agentic
and a top-down agentic approach. We assess code generated by the LLMs and
approaches in terms of \rev{compilability}, functional correctness, categories
of build errors, and the cost of translation in terms of the number of inference
tokens. Our results demonstrate that LLM translation of scientific applications
is feasible for small programs but difficulty with generating functional build
systems and cross-file dependencies pose challenges in scaling to larger
codebases.

%% file: intro.tex
High performance computing (HPC) hardware has increasingly diversified over the
last decade, now including GPUs from multiple vendors alongside CPU-based
systems. Portable programming models like Kokkos~\cite{kokkos:tpds2022} and
OpenMP~\cite{OpenMP4} provide a solution to develop a single source code for
multiple types of hardware, but challenges remain in productively \rev{converting}
existing code repositories to use them. The size and complexity of existing HPC
and scientific code repositories \rev{mean} that manually converting all relevant
portions of the codebase to use a new programming model is extremely
time-consuming, as well as potentially prone to error and the introduction of
performance issues.

Meanwhile, large language models (LLMs) have shown \rev{significant} promise in
automating simple programming tasks. While prior work has found that generating
parallel code for a task from scratch proves to be a challenge for current LLMs,
they are capable of translating existing serial or parallel code into a
particular parallel programming model~\cite{nichols:hpdc2024}. However, these
benchmark cases are translations of single, small functions or kernels.
Scientific applications are often composed of many such functions and kernels,
and solve more complex problems. Translating an entire software repository
introduces the complexities of data structure design, object hierarchies, and
coordination of work between multiple parallelized functions or kernels. If LLMs
can automate some of this work with minimal human intervention, the boon to
developer productivity would be immense. Nevertheless, whether or not LLMs are
capable of translating entire software repositories, including build systems,
headers, and multiple functions, between parallel programming models, remains an
open question.

In this paper, we attempt to answer the question posed above by proposing
\benchsuite, a suite of benchmarking cases for LLM-based automatic
\rev{repository-scale} program translation between parallel programming models.
This benchmark suite includes a range of standalone scientific computing and
artificial intelligence programs \rev{in} a range of codebase sizes, from
\rev{hundreds} to thousands of lines. The \rev{evaluation} tasks include
multiple programming model translations -- CUDA to OpenMP offloading, CUDA to
Kokkos, and OpenMP threading to OpenMP offloading. All but one of these cases
are chosen specifically to ensure that no \rev{publicly-}available translation
in the target programming model exists, to prevent the LLM from simply reciting
code memorized during training rather than reason through and solve an unseen
problem.

\benchsuite enables us to evaluate the capabilities of different LLMs in
correctly performing \rev{repository-scale} application translation, and enables
future researchers to design and test novel techniques for improved LLM-based
translation. We evaluate a non-agentic and a top-down agentic approach that can
use a variety of open-source and \rev{commercial} LLMs underneath, as well as
\rev{the} state-of-the-art SWE-agent~\cite{yang2024swe} agentic tool for
LLM-driven software engineering issue resolution. While some combinations of LLM
and translation techniques succeed in translation for smaller applications, we
demonstrate that \rev{repository-scale} application translation between portable
programming models poses unique challenges beyond the \rev{previously-observed
difficulty} with parallel code generation. Most importantly, regardless of the
translation technique or LLM used, creating working build systems
\rev{compatible with portable GPU programming models} and maintaining
consistency of interfaces across files remain major hurdles. Finally, we propose
a novel derived metric, expected token cost ($E_\kappa$), which predicts the
total token cost required to achieve a correct translation, and use it to
compare inference cost of translation techniques and LLMs.

In summary, this paper makes the following contributions:
\begin{itemize}
\item We design the \benchsuite benchmark for evaluating the ability of LLMs
  to translate entire software repositories between parallel programming models,
  which includes a range of application sizes and domains.
\item We evaluate the effectiveness of several state-of-the-art open- and
  closed-source LLMs using the \benchsuite benchmark, with agentic and
  non-agentic approaches of our own design, as well as a state-of-the-art LLM
  software engineering agent.
\item We propose a novel metric, $E_\kappa$, which measures the expected total
  inference cost in number of tokens required to generate a \rev{successful}
  translation, and use it to compare the inference costs of the translation
  techniques compared.
\item We study in detail the challenges LLM translation techniques encounter in
  generating working build system files (Makefiles, CMakeLists.txt, etc.) and
  handling cross-file dependencies, identifying which LLMs and translation
  techniques tend to make which specific mistakes.
\end{itemize}

%% file: bg.tex
In this section, we provide a brief introduction to large language models, their
use in code generation and translation, existing metrics for assessing generated
code, and agentic AI software engineering tools. We also include background on
the problems faced in manual translation of scientific applications between
parallel programming models. \rev{To avoid ambiguity we employ the term
\emph{LLM} to refer to large language models and \emph{programming model} to
refer to the parallel programming model (such as CUDA or Kokkos).}

\subsection{LLMs and reasoning LLMs}

\rev{LLMs have become the predominant approach to text generation. Given a
sequence of text, represented as a series of {\it tokens}, they iteratively
predict the next token in the sequence to generate an extended sequence. More
recently, LLMs have been fine-tuned with reasoning capabilities. These models
are trained to first generate reasoning to a solution before generating the
solution itself. DeepSeek-R1~\cite{deepseekai2025r1} and other efforts have
found that this reasoning training encourages the model to question and correct
itself. The result is LLMs with powerful reasoning capabilities that excel at
problem-solving tasks, including those that arise in programming.}

\subsection{LLMs for code generation}
\label{subsec:generation-bg}

\rev{The use of LLMs for code generation in high-performance computing contexts
has been widely
studied~\cite{nichols:hpdc2024,munley2023llm4vv,godoy2024large,valero2024chatblas}.}
As employed in prior efforts to assess the quality of parallel code generated by
LLMs~\cite{nichols:hpdc2024}, we adopt the pass@$k$ metric to estimate the
likelihood of getting a correct output given $k$ attempts by the LLM. We
\rev{define correctness} for our translation tasks in
Section~\ref{subsec:correctness}. Estimating pass@$k$ requires generating $N$
samples for a given LLM and task combination, where $N>k$. The number of correct
samples $c_t$ for a task $t$, as well as $N$ and $k$, determine the estimation
of pass@$k$ as demonstrated by \Cref{eq:pass-k}.

\vspace{1.5em}
\begin{equation}\label{eq:pass-k}
    \text{pass@}k =
    \frac{1}{\lvert \eqnmarkbox[MidnightBlue]{T1}{T}\rvert}
    \sum_{p\in \eqnmarkbox[MidnightBlue]{T2}{T}}
    \left[
        1 -
        \binom{
            \eqnmarkbox[WildStrawberry]{N1}{N} -
            \eqnmarkbox[OliveGreen]{ct}{c_t}
        }{
            k
        }
        /
        \binom{
            \eqnmarkbox[WildStrawberry]{N2}{N}
        }{
            k
        }
    \right]
\end{equation}
\annotatetwo[yshift=1em]{above}{N1}{N2}{Number of samples generated per task}
\annotatetwo[yshift=-1em]{below}{T1}{T2}{Set of tasks}
\annotate[yshift=-2.3em]{below,right}{ct}{Number of correct\\samples for task $t$}
\vspace{1.5em}

\subsection{Software engineering agents}
\label{subsec:ai-agent-bg}

\rev{There} has been growing interest in \emph{AI agents}: autonomous,
LLM-driven systems capable of creating or modifying code with minimal human
intervention. A notable example employed in this work is \emph{SWE-agent}, a
state-of-the-art coding agent designed to \rev{develop fixes for small software
engineering (SWE) issues. SWE-agent leverages LLMs in a closed-loop framework
including program analysis, generating and executing tests, and version control.
It first generates a high-level strategy for resolving a given issue, before
applying appropriate changes using its specialized editing tools, across
multiple files if needed. These features make SWE-agent particularly well-suited
for complex repository-scale refactoring.}

\subsection{Scientific application translation}
\label{subsec:manual-translation}

\rev{Translating scientific applications to new programming models is a
persistent challenge. As new GPU vendors and architectures proliferate, portable
GPU support has become essential. Many developers who previously ported to CUDA
now face the prospect of another porting effort to adopt portable models like
OpenMP or Kokkos~\cite{OpenMP4, kokkos:tpds2022}. Numerous case studies document
these transitions~\cite{malaya2023experiences,davis2023porting}. Regardless of
whether translation is manual or partially
automated~\cite{algarni2018framework,van2015integrated}, recurring themes
include the need for extensive user support, collaboration with vendors,
strategic planning, and considerable manual effort. Developers must consider
changes to build systems, function interfaces, parallelization strategies, and
algorithms. Converting scientific applications to portable GPU programming
models remains a difficult yet urgent task. In Section~\ref{sec:techniques}, we
present the methods we compare to address this challenge.}

%% file: techniques.tex
In this section we detail the techniques we benchmark for translation of HPC
software repositories.

\subsection{Non-agentic method}

The non-agentic strategy employs an LLM of the user's choice to translate a
repository file-by-file.
This approach is non-agentic, so no context or information can pass
between translations of separate files in the same repository. For
each file, we provide the contents of all other \rev{untranslated} files in
the repository as context. Listing~\ref{lst:na} provides a sample prompt, for the file
\verb|main.cpp| in nanoXOR.

\begin{lstlisting}[caption=Sample prompt for non-agentic translation method., label=lst:na]
You are a helpful coding assistant. You are helping a
software developer translate a codebase from the CUDA
execution model to the OpenMP Offload execution model.
Writing correct, fast code is important, so take some
time to think before responding to any query, and ensure
that the code you create is enclosed in triple backticks
(```), as used in the query below.

Below is a codebase written in the CUDA execution model.
We are translating it to the OpenMP Offload execution
model. Here is the file tree of the entire repository:

|-- Makefile
|-- README.md
+-- src/
    +-- main.cpp

Here is the code for each file in the codebase:

Makefile
...

src/main.cpp
...

Translate the src/main.cpp file to the OpenMP offload
execution model. Output the translated files in one code
block. Assume .cpp filenames whenever referring to other
files as this will be a C++ code.
\end{lstlisting}

\rev{To encourage accurate outputs, we begin with a system prompt including the
LLM's role, overall task, and emphasis that code should be correct and fast. We
provide the file tree for the full repository, and the full text of all the
untranslated files in the repository.} Finally, we indicate which file the LLM
should translate, the execution model to translate into, and the filename
extensions to assume where appropriate. For files containing main functions and
build system files (Makefile or CMakeLists.txt), we provide an addendum to
the prompt. For main function files, this addendum indicates the command line
interface the application should respect. For build system files, the addendum
indicates the interface the build system should respect, along with the compiler
\rev{and target architecture} it should be compatible with.

\subsection{Top-down agentic method}

\begin{figure}
  \centering
  \includegraphics[width=\linewidth]{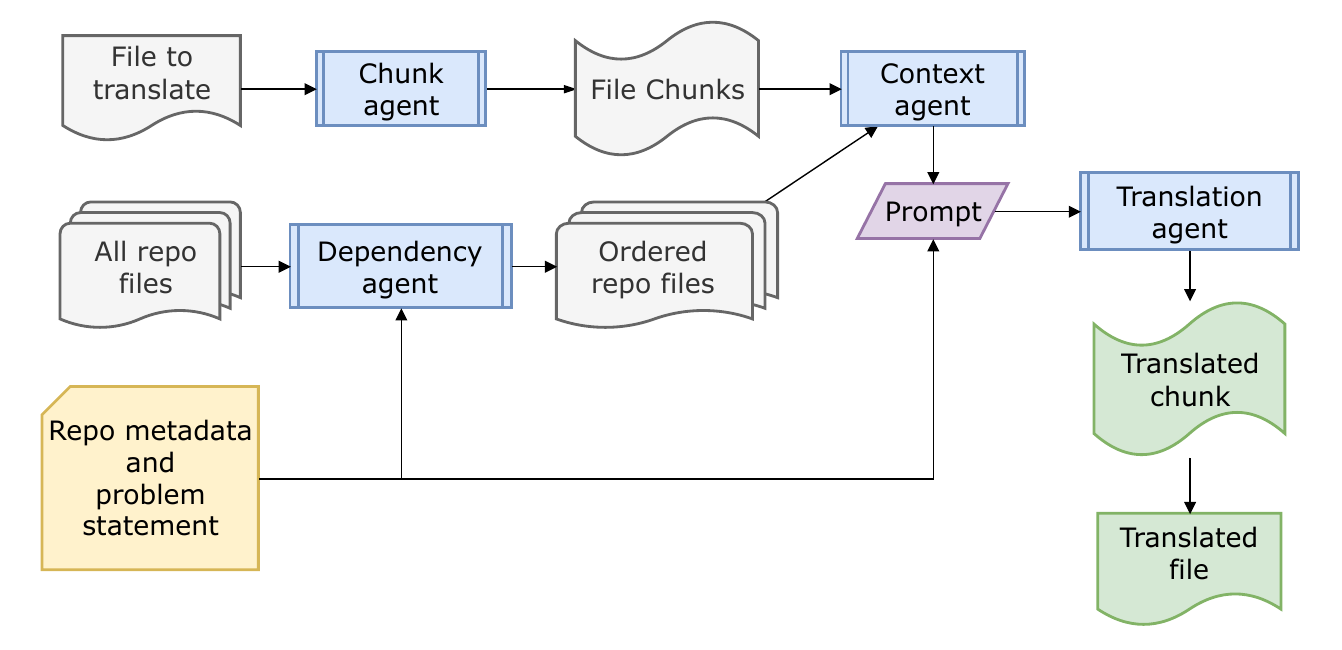}
  \caption{Control flow of the entire top-down agentic translation method.}
  \label{fig:tda}
\end{figure}

\rev{Directly translating an entire repository within the context window of an
LLM works well for smaller repositories, but will not scale to larger
repositories that do not fit in the context window. This means that the
translation will need to split into smaller translation tasks, working with
smaller portions of the full repository.} Finding the best way to partition the
data and work is non-trivial and various coding agents have been proposed that
address these issues with different approaches. We employ a top-down agentic
approach to translation, which is highlighted in Figure~\ref{fig:tda}.

Our top-down agentic approach is comprised of four LLM agents: chunk,
dependency, context, and translation. Each of the agents is a mix of static code
analysis and LLMs. The dependency agent runs first, determining the dependencies
between files in the repository. \rev{We translate files with no dependencies
first, since they do not require any external context. We utilize the clang
compiler to determine \texttt{\#include} dependencies only, precluding the
existence of circular dependencies.} For non-C/C++ files or C/C++ files where
clang fails, we use an LLM to analyze the file contents and repository structure
and determine the dependencies.

\rev{Files are then translated one by one in the order set by the dependency
agent. When translating files with dependencies we provide in the prompt a
summary of the changes already made to the dependencies. For example, if the
function \texttt{computeWithCuda()} is translated to
\texttt{computeWithOpenMP()}, then any files that call
\texttt{computeWithCuda()} need to be updated to call the new function. The
context agent produces LLM generated summaries of translation changes to pass
down to future dependents.}

\rev{The chunk agent is responsible for splitting files into smaller pieces that
fit within the context window of the translation agent's LLM. The chunk agent is
syntax-aware and splits files at the function level as much as possible, to
minimize scope splitting across chunks. Finally, the translation agent is an LLM
that translates code from one execution model to another.}

\subsection{SWE-agent}
\label{subsec:sweagent-technique}

We employ SWE-agent, described in Section~\ref{subsec:ai-agent-bg}, as a
technique for full-repository translation. To use SWE-agent for this
translation, we need to slightly reconfigure our translation task into a format
SWE-agent expects. SWE-agent addresses Github issues in git repositories, so we
start by rephrasing the usual translation prompt from the Non-agentic approach
as an issue, placed in a dedicated file that is passed to SWE-agent. We also
create a simple .git directory in the application codebase to be translated to
ensure SWE-agent sees the files it expects. Unfortunately, SWE-agent is designed
primarily to work with Python repositories, and automatically replaces tabs with
spaces, breaking Makefiles. This limits its usefulness for our benchmark suite.

%% file: llms.tex
In this section we briefly describe the large language models tested in this
study, including commercial, open-source, reasoning, and non-reasoning LLMs. To
keep total inference costs and node hours low, we use smaller and more
efficient versions of \rev{commercial} LLMs and quantized versions of open-source
LLMs.

\subsubsection*{Gemini 1.5 Flash}
Google AI released the multimodal Gemini 1.5 Flash in May 2024. Gemini model
parameter counts are not published~\cite{geminishort2023gemini}. We test
gemini-2.0-flash, but find performance to be severely worse compared to 1.5, so
those results are omitted. Gemini 1.5 Flash represents
state-of-the-art free but closed-source LLMs.

\subsubsection*{GPT-4o mini}
OpenAI's GPT-4o mini was released in July 2024. It is a smaller, lower-cost
version of GPT-4o, itself a multimodal variant of
GPT-4~\cite{openai2024gpt4ocard_short_author}. OpenAI models are closed-source
and model parameter counts are not published. We select GPT-4o mini as an
affordable paid LLM without reasoning capabilities.

\subsubsection*{o4 mini}
Another OpenAI offering, o4 mini is a more recent product from April 2025. It is
a smaller, more cost-efficient version of the o4 reasoning
model~\cite{openai2025o3_o4-mini_card}. Inference API costs for this model are
higher than GPT-4o mini. As a reasoning model o4 produces significantly more
output tokens per request.

\subsubsection*{Llama 3.3 70B Instruct}
Llama 3.3 is an open-source LLM developed by Meta
AI~\cite{dubey2024llama}, released in December 2024~\cite{llama3_3_model_card}.
We use the instruction fine-tuned version of
the model, a top performer in code tasks among open-source LLMs. To fit the 70
billion parameter model on one Delta node, we use a 4-bit GGUF quantization.
Llama 3.3 represents the state-of-the-art in open-source
non-reasoning LLMs.

\subsubsection*{QwQ-32B}
QwQ-32B is a 32-billion-parameter Qwen-based reasoning model
released in March 2025 by Alibaba Cloud~\cite{qwq_blog}. It performs
competitively versus other reasoning models like DeepSeek-R1, but
at smaller size. We select QwQ to represent state-of-the-art open-source
reasoning LLMs. We use an 8-bit GGUF quantization.

%% file: benchmarks.tex
In this section we describe the translation tasks we select for inclusion in
\benchsuite.\footnote{The full \benchsuite is available at \url{https://github.com/parallelcodefoundry/ParEval-Repo}}
\rev{Table~\ref{tab:benchmarks} summarizes all the benchmarking
tasks, including for each application the source lines of code (SLoC),
cyclomatic complexity (CC), number of files, and the programming models
translated between. Cyclomatic complexity serves as a general measure of
software complexity, specifically measuring the number of linearly independent
paths through the program. We collect it using the \texttt{pmccabe} tool.}
For all applications, we leverage the correctness validation test
cases provided by the developers to ensure the translation preserves
correctness.

\begin{table}[h]
  \centering
{ \small %
  \begin{tabular}{llll|llll}
    \toprule
                     & SLoC & CC  & \# Files & \rot{OMP Th.} & \rot{OMP Of.} & \rot{CUDA} & \rot{Kokkos}  \\
    nanoXOR          & 109  & 33  & 2        & \gchk         & \yque         & \gchk      & \yque   \\
    microXORh        & 127  & 33  & 3        & \gchk         & \yque         & \gchk      & \yque   \\
    microXOR         & 133  & 33  & 4        & \gchk         & \yque         & \gchk      & \yque   \\
    SimpleMOC-kernel & 780  & 59  & 6        &               & \yque         & \gchk      & \yque   \\
    XSBench*         & 2449 & 264 & 9        & \gchk         & \bcq          & \gchk      & \bcq    \\
    llm.c            & 3039 & 360 & 7        &               & \yque         & \gchk      & \yque   \\
    \bottomrule
  \end{tabular}
  \caption{The complete set of applications used as translation tasks in this
    work, including number of source lines of code (SLoC), cyclomatic complexity
    (CC), and number of files. On the right side, we indicate the programming
    models already implemented in the application in green and with a checkmark.
    Programming models we will attempt to port to are indicated with a question
    mark and colored yellow. For XSBench, we are porting to an existing model to
    examine whether the existence of a public port improves translation
    ability.}
    \label{tab:benchmarks}
}
\end{table}

\subsection{Applications selected}

The primary constraint on application selection is the existence of public ports
to the programming models we are attempting to translate into. If a public port
to that model does exist, data contamination, or the presence of the exact code
we want the LLM to generate in its training dataset, becomes likely. \rev{To
evaluate the impact of possible data contamination, we do include one case,
XSBench, which is known to have publicly-available ports in the programming
models we target.} As we will demonstrate in Section~\ref{sec:results},
applications at this level of complexity are sufficient to expose significant
weaknesses in current state-of-the-art approaches to full application
translation.

We briefly describe each application in \benchsuite{} below. Each application
differs in size and file count as well as key
characteristics that affect the difficulty of its associated translation tasks.

\subsubsection*{nanoXOR}

nanoXOR is a custom micro-application written specifically for \benchsuite{}. It
consists of a single kernel and driver function in a single source file. The
nanoXOR kernel performs a simple four-point stencil with the XOR operator over a
2D grid.

\subsubsection*{microXORh}
microXORh is also a custom micro-application, one step in complexity above
nanoXOR. It differs only in that the GPU kernel is written in a separate header
file that is included in the main function file, introducing a simple
compile-time dependency.

\subsubsection*{microXOR}
microXOR is the last of our custom micro-applications. It is one more step in
complexity up from microXORh, with the kernel and main driver function in two
separate source files, introducing a simple link-time dependency to the
translation problem.

\subsubsection*{SimpleMOC-kernel}
SimpleMOC-kernel is a proxy application for SimpleMOC~\cite{Tramm2016},
representing neutron flux attenuation. Only a CUDA version is available
publicly. It depends on the external cuRAND library, posing an
additional challenge to translation.

\subsubsection*{XSBench}
XSBench is a proxy application for OpenMC~\cite{tramm2014xsbench}, representing
macroscopic cross-section lookup. XSBench is a substantial step up in complexity
from SimpleMOC-kernel, but does have \rev{publicly}-available implementations of the
translations we attempt. This case will provide insight into whether possible
data \rev{contamination} can affect translation success rate.

\subsubsection*{llm.c}
llm.c is a CUDA implementation of LLM pretraining. We have slightly reduced the
size of the llm.c application to focus on critical application components. llm.c
provides a case study for making AI applications portable, and is a
more complex case that does not have \rev{publicly}-available ports to other
programming models.

\subsection{\rev{Programming model translation pairs selected}}

We further subdivide our translation tasks into pairs of programming models, the
source programming model and the destination programming model for translation.

\subsubsection{\rev{CUDA to OpenMP Offload}}

For all applications we examine translation from CUDA to OpenMP \rev{(GPU)
Offload. OpenMP compiler directives are relatively unobtrusive code
modifications, but compared to the much more obtrusive} and explicit CUDA
model, we should expect translations to require a significant degree of code
changes. Converting to OpenMP offload also poses \rev{compilation challenges,
as we require the LLM to produce a working Makefile that provides} correct
compile flags to use OpenMP offload for the specified compiler and system.

\subsubsection{\rev{CUDA to Kokkos}}

For all applications we also test translation from CUDA to
Kokkos~\cite{kokkos:tpds2022}, a portable programming model that can execute on
NVIDIA GPUs by acting as an abstraction layer over CUDA code. Kokkos code is
often structurally similar to CUDA code, including kernels and \rev{explicit
memory management. However, it poses greater challenge with successful
compilation, with the addition of a library dependency, advanced C++ features,
and the CMake build system generator.}

\subsubsection{\rev{OpenMP Threads to OpenMP Offload}}

For most applications, excluding those that lack a OpenMP \rev{threads (CPU)}
implementation, we evaluate translation from OpenMP threads to offload.
\rev{We would expect this task to be easiest, largely requiring modification of
existing OpenMP directives. However, this is the only translation task that
requires migration} from CPU to GPU parallelization.

In total, we evaluate translation of six applications for two translation pairs,
and four applications for the third translation pair, for a total of sixteen
unique translation tasks.

%% file: metrics.tex
In this section, we define the metrics we use to assess the quality of
translations of full-scale applications between parallel programming models.
We consider the quality of the translation tool output in terms of both
correctness and token economy.

\subsection{Metrics for correctness}
\label{subsec:correctness}

For correctness, we adopt the definition of pass@$k$ from
~\cite{nichols:hpdc2024}, as described in~\ref{subsec:generation-bg}
by \Cref{eq:pass-k}. We report both the average of pass@$k$ over all tasks in
\benchsuite{} as well as per-task pass@$k$.

Each application in our benchmark suite provides test cases, and a
solution must generate expected outputs \rev{for
those tests. To be considered correct we also require that the translation be
implemented using the requested target programming model and execute on the
hardware specified in the prompt.}

\rev{To separately consider compilation success and failure this in our
evaluation, we employ build@$k$, which measures the probability that the model
will generate a compilable solution given $k$ attempts. This metric is similar
to pass@$k$ except that it considers all samples that compile, not just those
which are correct. The number of correct samples, $c_t$, is replaced with the
number of buildable samples, $b_t$. The typical translation task for a full
application can take some time, on the order of ten to thirty minutes. As such,
$pass@1$ and $build@1$, or the likelihood of correctness and build success for
a single translation attempt, provide the best insight into user experience
among possible $k$ values.}

\subsection{Metric for token economy}
\label{subsec:token-economy}

We also measure the token economy of LLM
inference prompts generated by the translation tool, to understand how tool
choice affects cost of translation. Most LLMs inference serving
APIs charge users by the number of input tokens and output tokens used, and
output tokens are typically more expensive than input tokens. \rev{Self-hosted
LLM inference does not directly incur a financial cost, but does incur
a compute time cost in node hours, charged to an allocation if using a
cloud or supercomputing center. In such cases, the number of tokens consumed is
proportional to the compute time of the inference prompt.} We report for each
combination of translation tool and translation task the average number of
tokens consumed. \rev{The tokenization process differs across LLMs},
so these values should be compared primarily between different LLM-based tools
sending prompts to the same underlying LLM. \rev{To enable comparison between
LLMs, we provide estimated costs in dollars or node hours as appropriate in
Section~\ref{subsec:economy-results} for two top-performing LLMs.}

To assess the \rev{impact of} accuracy on token economy, we propose the
\textit{expected token cost} ($E_\kappa$) metric. This is defined as the
expected number of generations required to produce a correct translation (i.e.,
the multiplicative inverse of the pass@1 score) multiplied by the average number
of tokens required to produce a single translation. $E_\kappa$ is defined
explicitly in Eq.~\ref{eq:expect}.

\vspace{1.5em}
\begin{equation}\label{eq:expect}
    E_\kappa =
    \left(
    \frac{1}{\eqnmarkbox[MidnightBlue]{P1}{\text{pass@}1}}
    \right)
    \cdot
    \eqnmarkbox[WildStrawberry]{k}{\kappa}
\end{equation}
\annotate[yshift=-1em]{below}{P1}{Single generation correctness chance}
\annotate[yshift=1em]{above}{k}{Average token cost per\\generation for the task}
\vspace{1.5em}

\subsection{Identifying and classifying translation errors}
\label{subsec:errors}

\rev{To identify specific opportunities for improvement, we analyze mistakes
made by LLMs in their translation attempts. Failures in the translation
process can come from many sources and are only recorded in unstructured build
and run logs. To evaluate errors from the thousands of builds and runs we
perform, we develop a semi-automated method to classify errors in the
translation process. We cluster the build and run logs of the translated
applications into sets of similar errors, and then name these clusters based on
the error class they reflect.}

\rev{We first convert the build and run logs of each
translation and run to vector embeddings using the
\texttt{word2vec} \cite{word2vec} model. This yields for each translation a
single vector that captures the semantics of its output logs. We then cluster
these vectors using the \texttt{DBSCAN}~\cite{dbscan} algorithm, a
density-based clustering algorithm that can identify clusters of arbitrary
shapes, is robust to noise, and has only two hyperparameters that need to be
tuned.} We manually inspect the quality of resulting clusters to tune
DBSCAN's hyperparameters.

While the clustering is helpful, it produces \rev{many clusters and fails to
join some similar errors. We make a manual pass over the
algorithmically-generated clusters to merge them and reassign samples to
different clusters where appropriate. During the manual pass we also
assign a short label to each cluster that describes the category of error it
contains.} This combination of automated clustering, followed by manual
correction and labeling, allows us to productively classify the most common
errors in the translation process for a large number of translation tasks.

%% file: setup.tex
In this section we describe the experimental setup for translation generation
(inference) and testing.

\subsection{Inference setup}

For commercial API models, we set a total budget of 200 dollars, with 170
dollars for o4-mini and 30 dollars for GPT-4o-mini. We complete Gemini
inference using the Google's API free tier.

For locally-hosted open-source models, we employ the vLLM inference engine
version 0.8.2 configured in server mode. We run inference on the Delta system at
NCSA\footnote{\url{https://docs.ncsa.illinois.edu/systems/delta/en/latest/index.html}},
specifically the single-socket A100 nodes, which have four 40 GB NVIDIA A100
GPUs, one 64-core AMD Milan CPU, and 256 GB of RAM. We configure vLLM to use all
four GPUs with tensor parallelism with prompt prefix caching enabled, and batch
requests where possible.

\subsection{Evaluation setup}

All translation tasks in \benchsuite{} require translation to a GPU
programming model. We test translations on a single NVIDIA 40GB A100
GPU, hosted on the Zaratan system at the University of
Maryland\footnote{\url{http://hpcc.umd.edu}} on a single node with a
128-core AMD EPYC 7763 CPU and 512 GB of RAM. We use CUDA 12.3.0
universally and compile OpenMP offload codes with LLVM 19.1.7 and
Kokkos codes with Kokkos 4.5.01 and GCC 11.3.0.

%% file: results.tex
In this section we present \rev{example translations,} results for correctness
of generated code translations across LLMs, tasks, and techniques, our analysis
of the clustering of errors encountered in attempting to compile LLM
translations, and our analysis of the token \rev{efficiency} of cases where the
LLM translation is \rev{successful}.

\subsection{\rev{Examples of successful and unsuccessful translations}}
\label{subsec:translation-examples}

\rev{Listings~\ref{lst:cuda}-~\ref{lst:omp-ag} provide the original nanoXOR CUDA
kernel, and examples of correct and incorrect translations to OpenMP Offload
generated by QwQ with the non-agentic and agentic translation techniques,
respectively. The agentic translation is incorrect, as it omits the \texttt{target}
and \texttt{parallel for} directives, which are required to run the loop on the GPU
and parallelize the loop across threads.}

\begin{lstlisting}[caption=\rev{Original nanoXOR CUDA kernel.}, label=lst:cuda, language=C++]
__global__ void cellsXOR(const int *input, int *output, size_t N) {
  int i = blockIdx.y * blockDim.y + threadIdx.y;
  int j = blockIdx.x * blockDim.x + threadIdx.x;
  if (i < N && j < N) {
    int count = 0;
    if (i > 0 && input[(i-1)*N + j] == 1) count++;
    if (i < N-1 && input[(i+1)*N + j] == 1) count++;
    if (j > 0 && input[i*N + (j-1)] == 1) count++;
    if (j < N-1 && input[i*N + (j+1)] == 1) count++;
    output[i*N + j] = (count == 1) ? 1 : 0;
  }
}
\end{lstlisting}

\begin{lstlisting}[caption=\rev{Correct nanoXOR OpenMP Offload translation generated by the non-agentic translation method with \texttt{qwq-32b-q8\_0}.}, label=lst:omp-nag, language=C++]
#pragma omp target data map(to: input[0:N*N]) map(from: output[0:N*N])
{
  #pragma omp target teams distribute parallel for collapse(2)
  for (int i = 0; i < N; i++) {
    for (int j = 0; j < N; j++) {
      // Kernel body identical, omitted for brevity
\end{lstlisting}

\begin{lstlisting}[caption=\rev{Incorrect nanoXOR OpenMP Offload translation generated by the agentic translation method with \texttt{qwq-32b-q8\_0}.}, label=lst:omp-ag, language=C++]
#pragma omp target data map(to: input[0:N*N]) map(from: output[0:N*N])
{
  #pragma omp teams distribute collapse(2) num_threads(blockEdge * blockEdge)
  for (size_t i = 0; i < N; i++) {
    for (size_t j = 0; j < N; j++) {
      // Kernel body identical, omitted for brevity
\end{lstlisting}

\subsection{Translation correctness}
\label{subsec:correctness-results}

\begin{figure*}[h]
  \centering
  \begin{subfigure}{0.5\textwidth}
    \centering
    \includegraphics[width=0.71\linewidth]{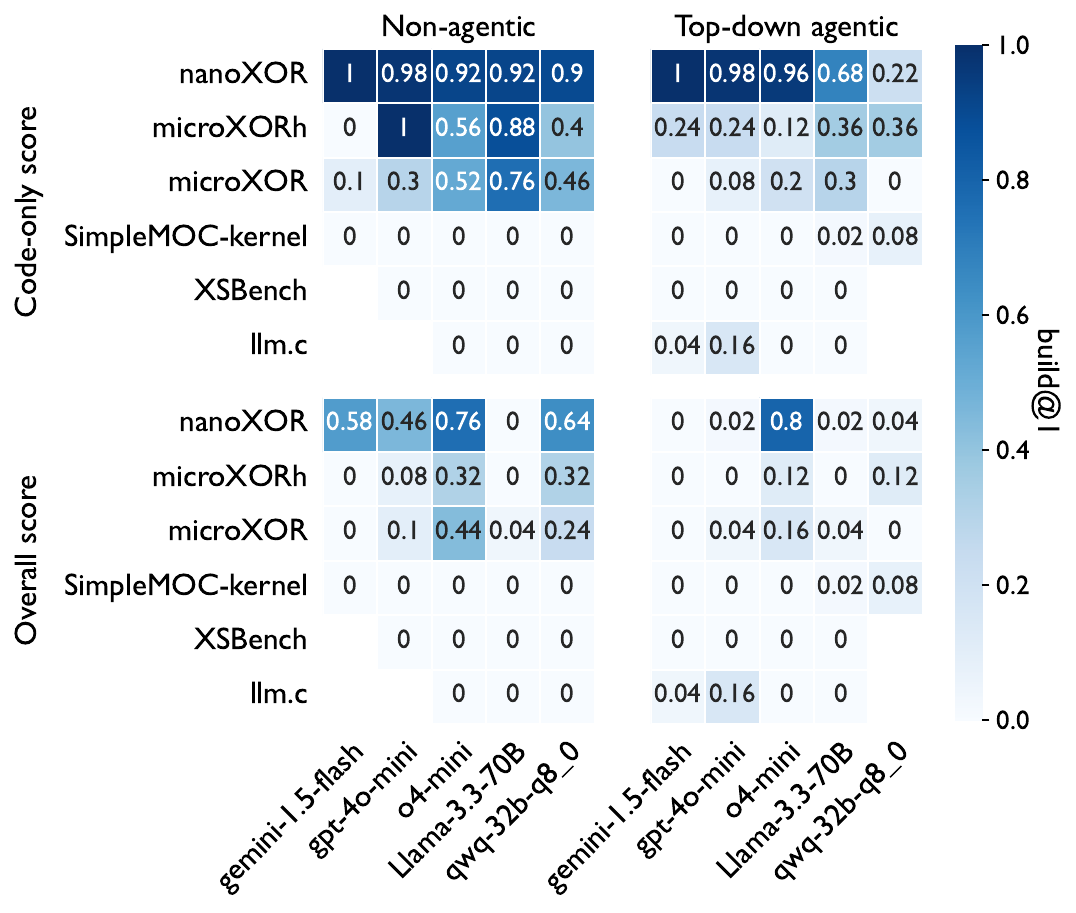}
    \caption{build@1 for CUDA to OpenMP Offload}
    \label{fig:cuda-omp-build}
  \end{subfigure}%
  \begin{subfigure}{0.5\textwidth}
    \centering
    \includegraphics[width=0.71\linewidth]{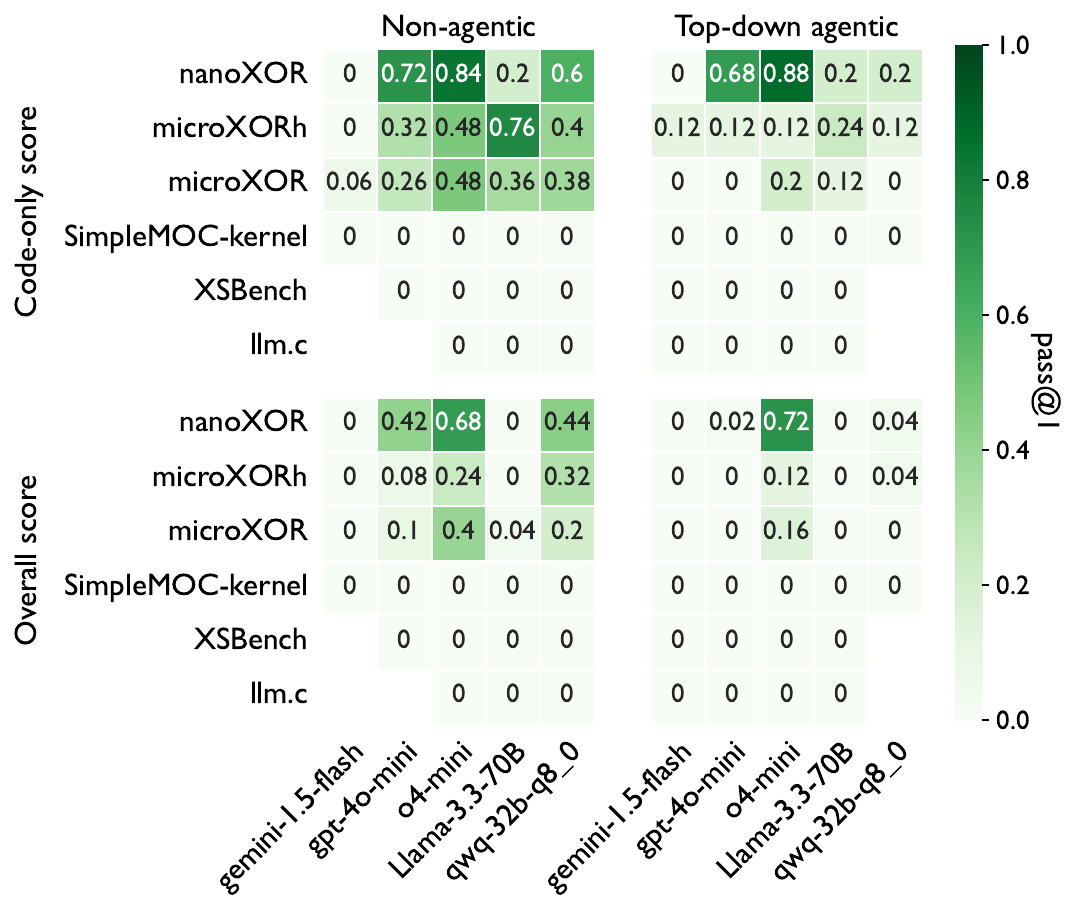}
    \caption{pass@1 for CUDA to OpenMP Offload}
    \label{fig:cuda-omp-pass}
  \end{subfigure}

  \begin{subfigure}{0.5\textwidth}
    \centering
    \includegraphics[width=0.99\linewidth]{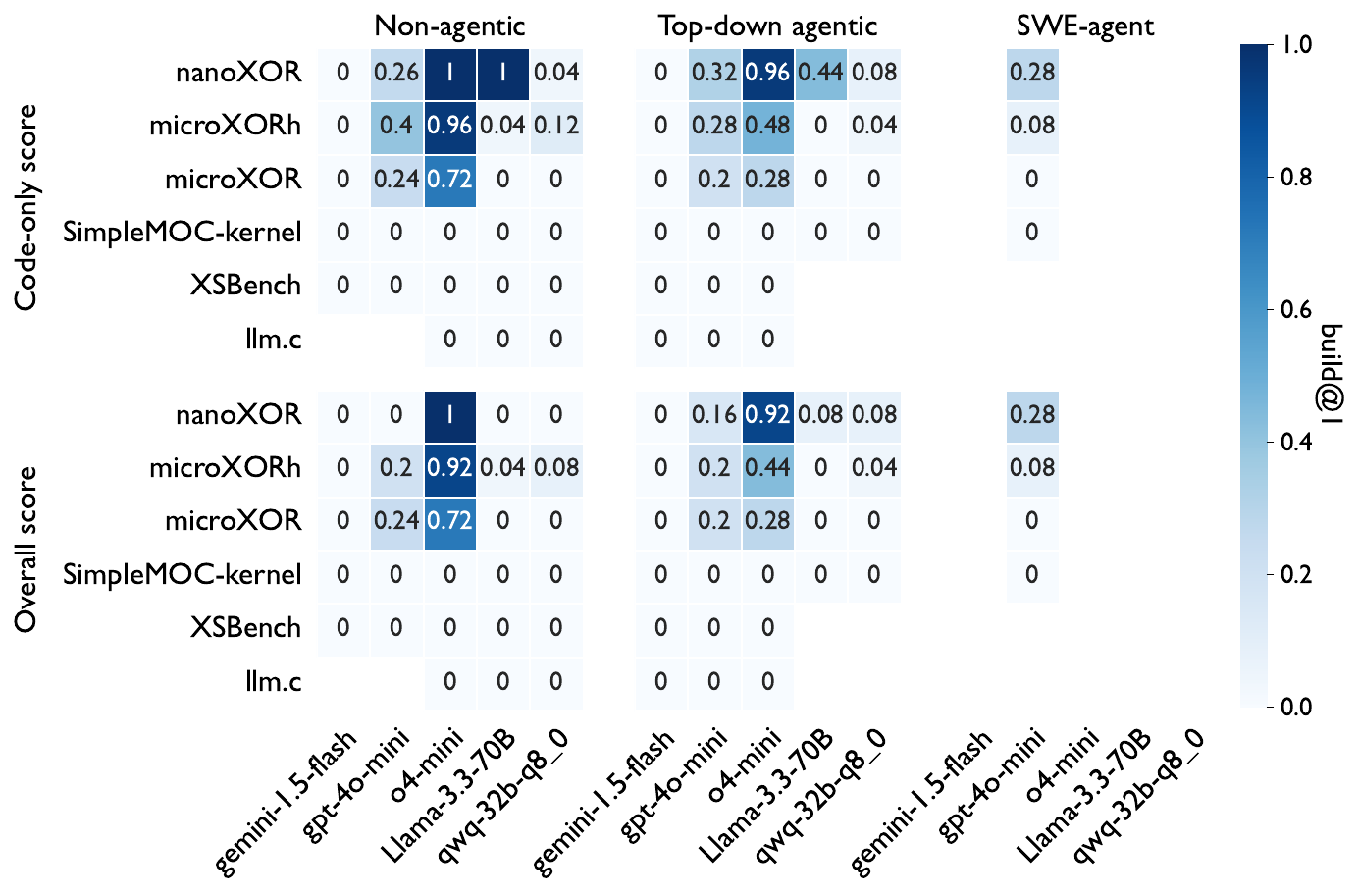}
    \caption{build@1 for CUDA to Kokkos}
    \label{fig:cuda-kokkos-build}
  \end{subfigure}%
  \begin{subfigure}{0.5\textwidth}
    \centering
    \includegraphics[width=0.99\linewidth]{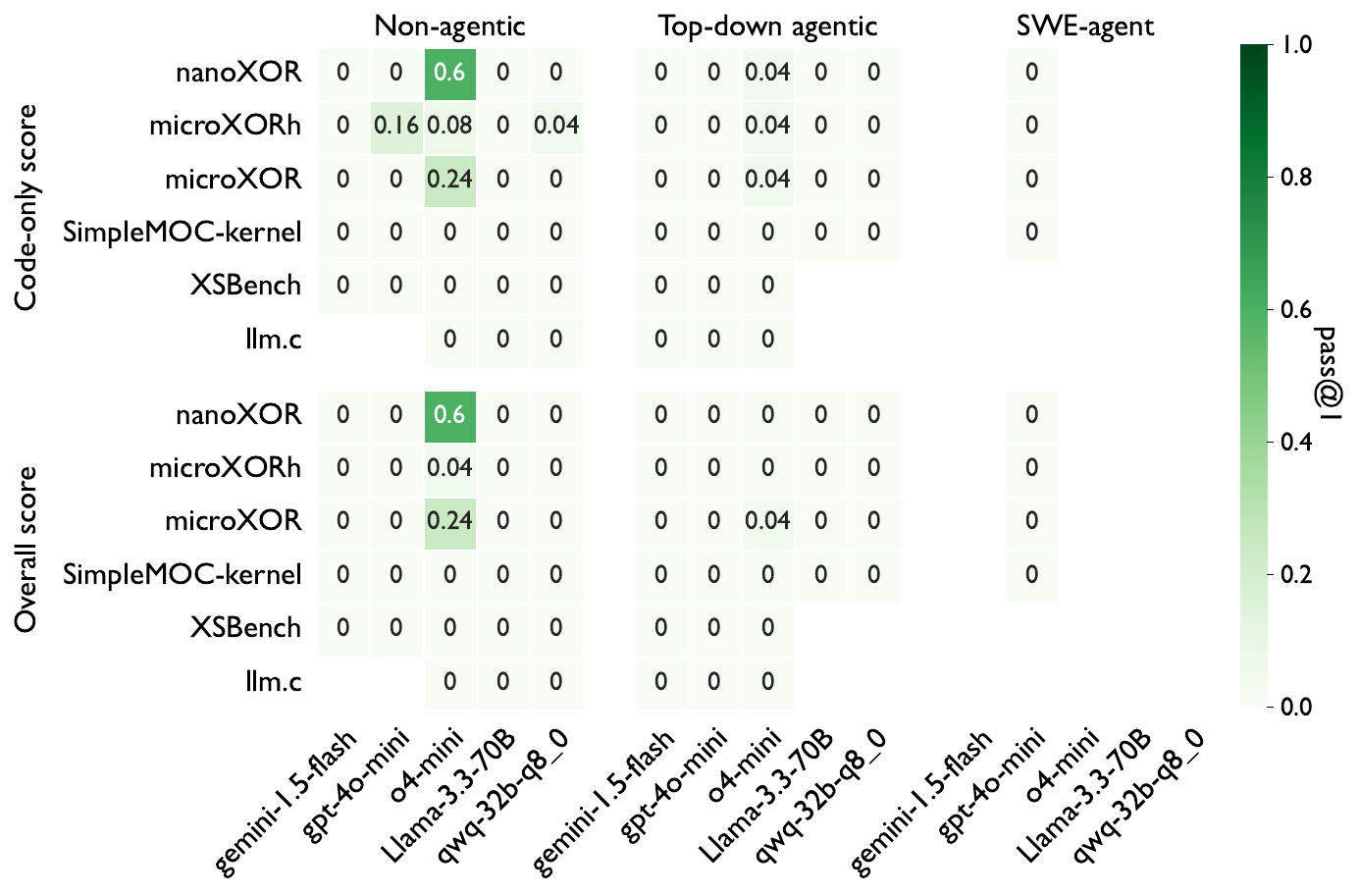}
    \caption{pass@1 for CUDA to Kokkos}
    \label{fig:cuda-kokkos-pass}
  \end{subfigure}

  \begin{subfigure}{0.5\textwidth}
    \centering
    \includegraphics[width=0.66\linewidth]{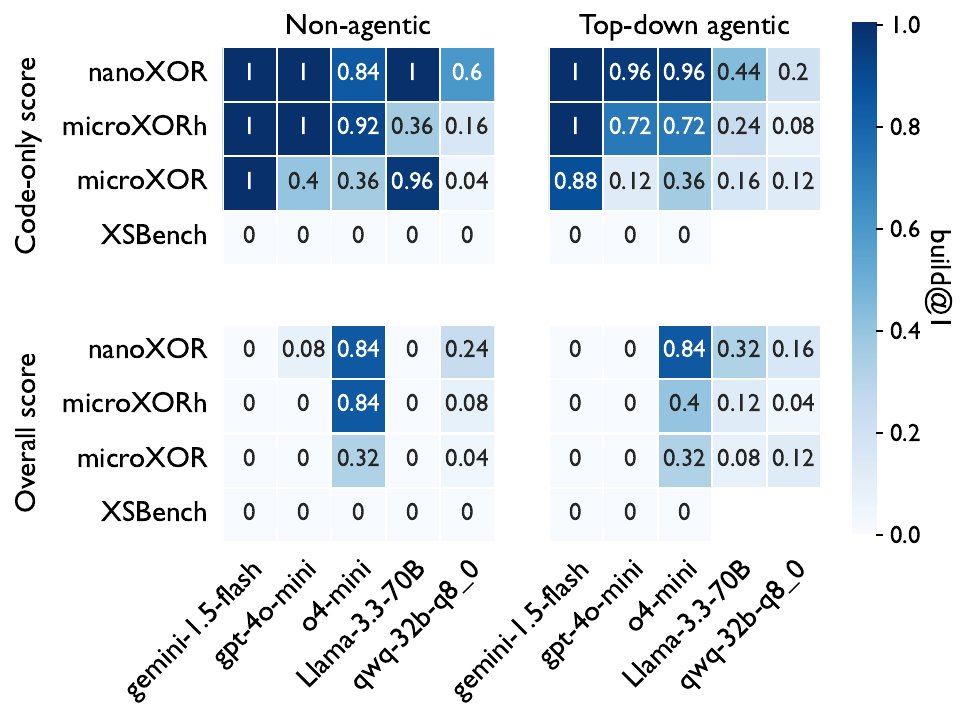}
    \caption{build@1 for OpenMP Threads to OpenMP Offload}
    \label{fig:omp-omp-build}
  \end{subfigure}%
  \begin{subfigure}{0.5\textwidth}
    \centering
    \includegraphics[width=0.66\linewidth]{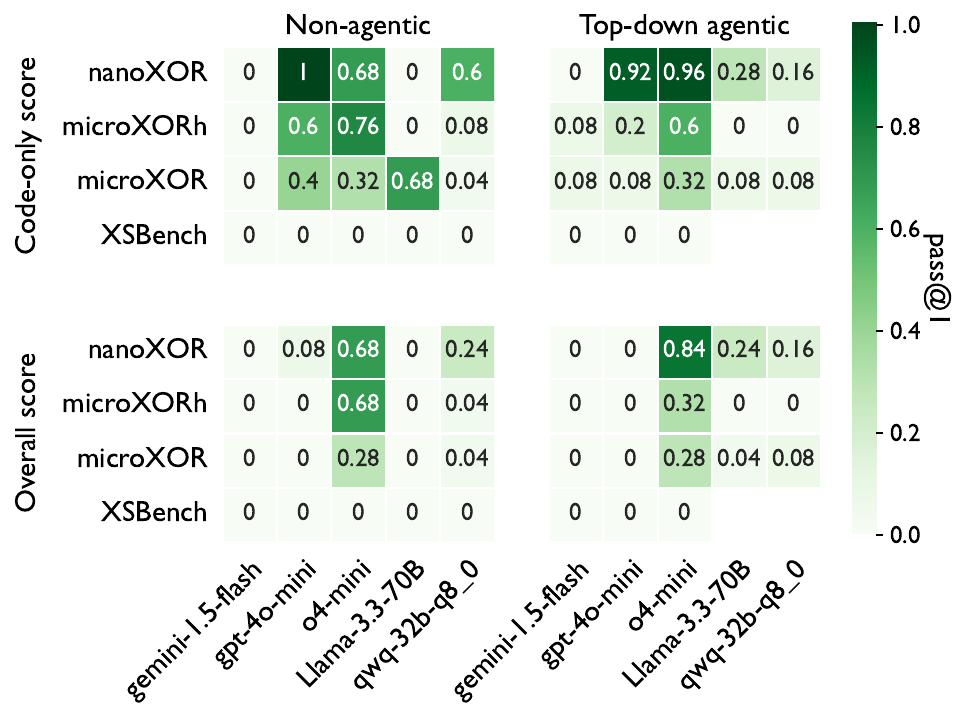}
    \caption{pass@1 for OpenMP Threads to OpenMP Offload}
    \label{fig:omp-omp-pass}
  \end{subfigure}
  \caption{Correctness metrics for OpenMP Threads to OpenMP Offload tasks.}
  \label{fig:correctness}
\end{figure*}

In Figure~\ref{fig:correctness} we present build@1 and pass@1 results for our
three programming model translation pairs. Within each row of subfigures, the
left subfigure indicates build@1 scores, the likelihood of generating a
compilable translation given one attempt, while the right subfigure indicates
pass@1, the likelihood of generating a translation that passes correctness tests
given one attempt.

Within each subfigure, the heatmaps are organized into rows and columns. Each
column of heatmaps corresponds to a translation technique, non-agentic,
top-down, and SWE-agent (where applicable). The upper row of heatmaps in each
subfigure corresponds to ``Code-only'' score, while the lower row of heatmaps
corresponds to ``Overall'' score. ``Code-only'' score considers only the
correctness of the generated source code, using a pre-written ground truth
Makefile or CMakeLists.txt manually translated by the authors to compile the
LLM-translated source code. ``Overall'' score reflects the results of using
the LLM-translated source code and build system.

\rev{Note that an experiment configuration with no value in its
heatmap cell indicates that we do not run that case, and results of 0 indicate
that we run the configuration but it does not generate any correct translations.}
In several cases, we are unable to generate translations for a combination of
LLM, translation technique, application, and programming model pair. The
non-agentic approach, because it requires each file to be generated in one response,
cannot scale to some application sizes due to
exceeding LLM output context limits. This is the case for Gemini and GPT-4o when
translating llm.c, and for Gemini when translating XSBench from CUDA to OpenMP
offload. In some cases with the top-down agentic approach using local models, we
do not complete translation due to exceeding our per-experiment budget of 8
node hours. This is the case for \rev{QwQ} translating XSBench and llm.c with
all programming model pairs as well as Llama-3.3 translating XSBench and llm.c
from CUDA to Kokkos. Finally, we present SWE-agent results for a subset of the
cases due primarily to the SWE-agent's incompatibility with Makefile mentioned
in Sec.~\ref{subsec:sweagent-technique}, but also omit XSBench and llm.c to
remain within our OpenAI API budget.

Across all cases, we observe that Overall score is consistently \rev{significantly}
lower than Code-only score. This indicates a substantial \rev{capability} gap between
source code translation and build system generation. To further explore the
reasons why LLMs fail to generate functional build systems, we examine the types
of build errors encountered in this study in Sec.~\ref{subsec:error-results}.

Examining the impact of translation technique, we observe that
the non-agentic approach, where it can complete a translation, tends to achieve
the highest scores in both build and pass@1, although this varies across LLMs
used. SWE-agent, while only tested with one LLM and programming model pair,
achieves moderate success, particularly in overall build@1 score. But, it does not
manage to generate any code that passes correctness tests. The non-agentic
likely outperforms the top-down agentic approach due to the greater quantity of
repository context provided, highlighting the need for more sophisticated
approaches in future work to develop full application translation
agents. Along the programming model translation pair axis, we observe that CUDA
to Kokkos translation is significantly more challenging that the translations
involving OpenMP.

Turning next to the application axis, we find that more complex applications
generally pose greater challenge for all LLMs and translation techniques, as
expected. One notable exception can be found with non-agentic Llama-3.3
translating CUDA to OpenMP offload, where code-only pass@k is 0.76 for microXORh
but only 0.2 for nanoXOR. Additionally, non-agentic Llama-3.3 translating OpenMP
threads to OpenMP offload achieves a pass@k of 0.68 for microXOR and 0 for
microXORh and nanoXOR. Overall, no combination of translation technique and LLM
achieves a pass@k above 0 for any application larger than microXOR. This key
finding indicates that LLM-based translation cannot yet produce working code in
a fully automated manner. In Sec.~\ref{subsec:error-results}, we explore the key
obstacles to success in translating larger applications in terms of most
frequent compilation errors in translated code.

\subsection{Error clustering}
\label{subsec:error-results}

\begin{figure}[h]
  \centering
  \includegraphics[width=\linewidth]{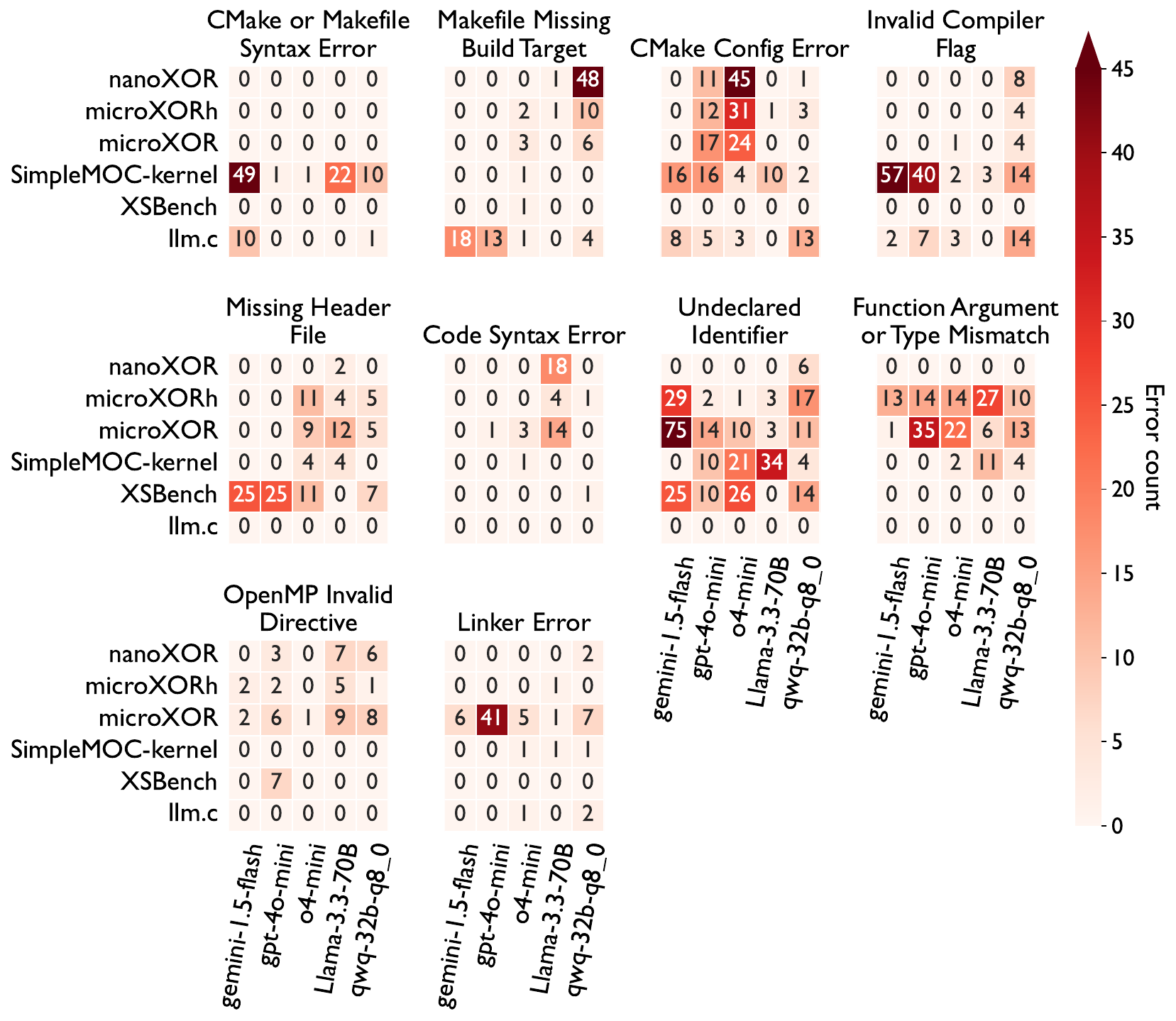}
  \caption{Count of categories of errors encountered when across combinations of large language models and application.}
  \label{fig:errors}
\end{figure}

We cluster the error messages recorded by
\benchsuite{}'s translation testing code when attempting to build the generated
translations, as described in Sec.~\ref{subsec:errors}. Figure~\ref{fig:errors}
displays the results of this analysis, after manually combining highly similar
clusters and removing clusters of less interest, including errors related to
missing files and build timeouts as well as successful build outputs. Across
applications and LLMs, errors relating to CMake configuration, as well as
undeclared identifiers and function argument or type mismatches for apps besides
nanoXOR, are broadly common. This suggests that coordination of function
interfaces and variable names across files and successful configuration of CMake
for Kokkos builds are common points of difficulty across LLMs. These broader
issues must be addressed by improved LLM translation techniques and prompting,
rather than tuning the choice of LLM.

However, we also observe that some categories of errors are only highly
prevalent for some LLMs and applications. For example, Gemini appears most
likely to struggle with Makefile syntax and compiler flags, particularly for
SimpleMOC-kernel, Llama-3.3 is particularly \rev{susceptible} to source code
syntax mistakes, and GPT-4o mini produces translation that fail to link
especially often for microXOR. \rev{We observe that LLMs can still encounter
unique pitfalls that may be avoided by tuning the choice of LLM.}

\rev{While many of these error categories can occur for ordinary, non-HPC codes,
we note that the use of portable GPU programming models introduces unique
complexities in compilation that contributes significantly to their prevalence,
as observed in prior work~\cite{davis:ics2025}. For example, mistakes with
compiler flag choice frequently arise from use of incorrect OpenMP Offload
flags, and the undeclared identifier and function argument/type mismatch
categories include mistakes in interacting with Kokkos library features.}

\subsection{Inference token economy}
\label{subsec:economy-results}

\begin{figure}[h]
  \centering
  \includegraphics[width=\linewidth]{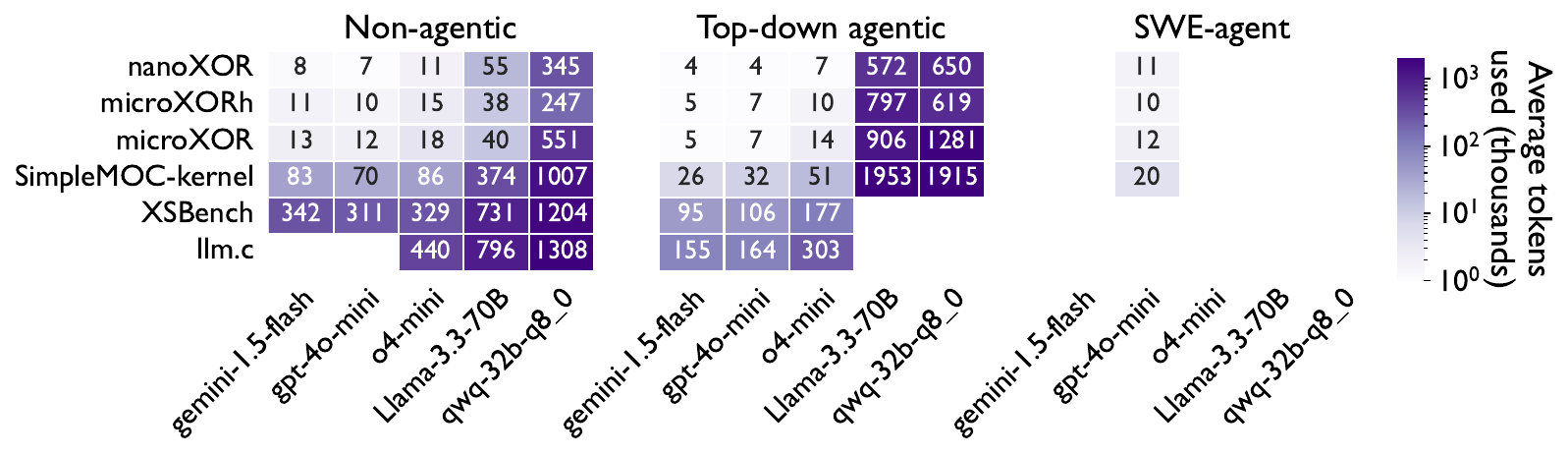}
  \caption{Total inference tokens used in translation, averaged across
    individual generations and programming model translation pairs.}
  \label{fig:tokens}
\end{figure}

We also analyze the cost of translation in terms of token economy, or the number
of inference tokens required to complete a translation. Figure~\ref{fig:tokens}
displays total inference tokens used for translation for each technique, LLM,
and application combination, averaged across programming model translation pairs
and individual generations. \rev{Note that in Figure~\ref{fig:tokens} some
additional heatmap cells are empty compared to Figure~\ref{fig:correctness},
because we do not include the total tokens consumed for any cases where zero
total correct translations are generated.} Among \rev{commercial} API LLMs, the
non-agentic translation technique consumes more inference tokens than top-down,
but among open-source locally-hosted LLMs, the top-down agentic approach is more
expensive, largely because the commercial API LLMs are more conservative in
selecting translation context in the top-down approach. In the non-agentic
approach, QwQ in particular consumes a significant quantity of tokens, due to
the size of its reasoning output. o4-mini is also a reasoning model but consumes
significantly fewer additional tokens.

\begin{figure}
  \centering
  \includegraphics[width=0.8\linewidth]{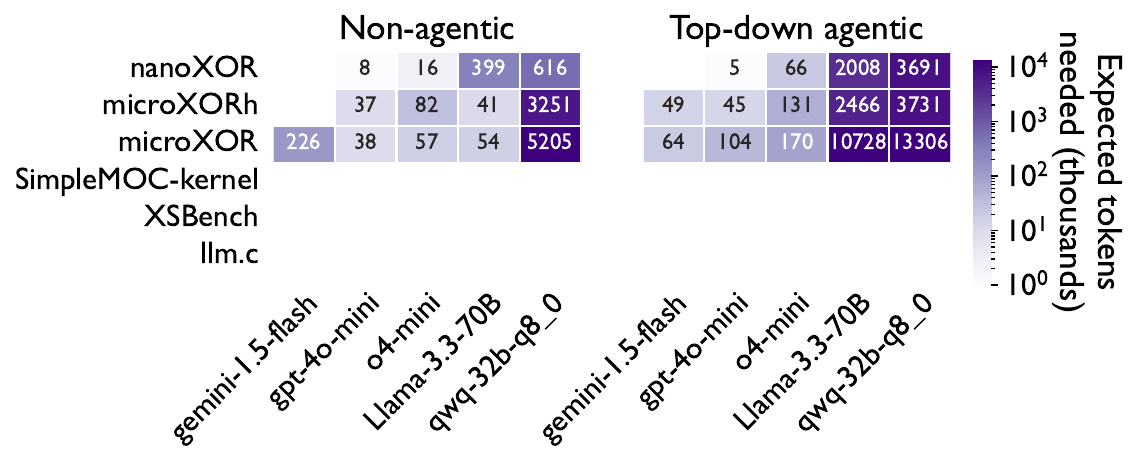}
  \caption{Expected tokens needed for successful translation ($E_\kappa$),
    averaged across individual generations and programming model translation
    pairs where at least one generation passed correctness \rev{test is} (i.e., pass@1
    is greater than 0).}
  \label{fig:expect}
\end{figure}

Figure~\ref{fig:expect} lists $E_\kappa$, the expected number of tokens needed
to produce a \rev{successful} translation, as defined in
Sec.~\ref{subsec:token-economy}. We aggregate this metric only over cases where
the pass@1 is greater than 0. We can conclude from Figure~\ref{fig:expect} that
among \rev{commercial} API models, non-agentic o4-mini produces correct translations
at the lowest token cost, while among open-source locally-hosted models,
non-agentic Llama-3.3 is cheapest.

Given the expected token cost provided in Figure~\ref{fig:expect}, we can estimate
the total cost in US dollars or in node-hours for the least expensive API and
open-source models, respectively. Table~\ref{tab:cost-est} lists these estimates
for the three applications that can be successfully translated. We calculate these
estiamtes using public OpenAI API costs for o4-mini as well as our
observed average generation throughput of 187 tokens per second on a single node
of the Delta system with vLLM, as described in~\ref{sec:setup}. Note that
Llama-3.3 nanoXOR cost is particularly high due to Llama's unexpected difficulty
with that simple application, as described in
Sec.~\ref{subsec:correctness-results}.

\begin{table}[h]
  \centering
{ \small %
  \begin{tabular}{l|lll}
    \toprule
                          & nanoXOR  & microXORh & microXOR \\
    Non-agentic o4-mini   & \$0.03   & \$0.04    & \$0.05 \\
    Non-agentic Llama-3.3 & 0.6 n.h. & 0.06 n.h. & 0.08 n.h. \\
    \bottomrule
  \end{tabular}
  \caption{Estimated cost in dollars or node-hours for successful translation,
    based on $E_\kappa$, published OpenAI API costs, and our observed average
    generation throughput on a single Delta node with Llama-3.3 in vLLM (187
    tokens per second).}
    \label{tab:cost-est}
}
\end{table}

%% file: related.tex
Prior studies have examined using LLMs for code translation. We separate the
relevant literature into two categories, repo-level translation and parallel
code translation and generation.

\subsection{Repository-level code translation}

Automating repository-level translation has been a growing area of focus. Prior
studies leverage LLMs to minimize human intervention in the translation process
when scaling translation to larger codebases. For instance, AlphaTrans explores
translating Java to Python by breaking down large repositories into smaller,
manageable fragments and iteratively validating results for syntactic
correctness and functional equivalence~\cite{ibrahimzada2024repository}.
Additional work also considers the problem from a benchmarking perspective.
RepoTransBench proposes a benchmark suite of Python to Java translation tasks
for assessing LLM translation capabilities, finding that LLMs \rev{consistently}
achieve below-par results for these tasks~\cite{wang2024repotransbench}.
However, these projects both consider translation between Java and Python,
without parallel programming models and with no consideration of languages more
common in high-performance computing, including C/C++ and Fortran. Furthermore,
the methodologies in these studies is tightly integrated with Java and Python
code and build systems, making the techniques difficult to generalize to other
workflows, particularly those in HPC. As an example, we discovered that
SWE-agent does not work with Makefiles, which are a critical part of many HPC
codes.

\subsection{Parallel code translation}

Several efforts to develop LLM tools to translate parallel code exist. First,
CodeRosetta~\cite{tehrani2024coderosetta} proposes an encoder-decoder
transformer model to translate between C++ and CUDA as well as Fortran and C++.
Unfortunately, its approach is evaluated only using the CodeBLEU similar score,
without genuine runtime data. Similarly, Code-Scribe utilizes a scoped code
hierarchy and retrieval-augmented generation (RAG) to guide LLM-based
translations from Fortran to C++, ensuring compatibility through intermediate
Fortran-C APIs~\cite{dhruv2024leveraging}. LASSI, on the other hand, focuses on
translating parallel programming models, such as CUDA and OpenMP, using
iterative compilation and runtime feedback for
self-correction~\cite{dearing2024lassi}. Similarly, the dataset introduced in by
Lei et al.~\cite{lei2023creatingdatasethighperformancecomputing} focuses on CPU
OpenMP Fortran and C++. but does not extend or generalize to other programming
models like Kokkos or OpenMP GPU offloading, and relies on only the CodeBLEU
similar metric and human evaluation.

While these strategies demonstrate progress in translating parallel code, they
often do not consider genuine code functionality, instead employing CodeBLEU
code \rev{similarity} scores to assess translation quality. Furthermore,
many employ translation tasks that have already have \rev{publicly}-available
implementations. For example, LASSI relies on HeCBench, which include reference
implementations for CUDA and OpenMP already. This creates a risk of training
dataset contamination with test problems, making it possible for LLMs to recite
already-seen translated code rather than reasoning through the problem and
carrying out a genuine translation.

%% file: conc.tex
Converting an entire application codebase to use a new programming model is a
tedious but necessary task to take advantage of the hardware available on
flagship supercomputing platforms. Leveraging large language models to automate
this effort has the potential to enormously improve HPC application developer
productivity. In this paper, we have described \benchsuite, a collection of HPC
application translation tasks covering multiple parallel programming models and
a range of application sizes. We have evaluated a range of state-of-the-art LLMs
using both non-agentic and agentic approaches to full application translations,
and find that existing techniques largely fail to translate applications beyond
trivial scale. We identify that generating working build systems is a major
obstacle to successful full-repository translation, and furthermore identify
categories of compilation failures that occur across LLMs and only for specific
LLMs. The insights gained from our analysis will be critical
for designing future approaches that can successfully translate large
applications. We conclude with a cost analysis of the most token-economic
open-source and \rev{commercial} LLMs for our translation tasks using our
proposed metric for expected token cost, $E_\kappa$. \rev{Given the difficulty
LLMs encounter in generating compilable translations, opportunities for future
work include constructing a dataset of complete HPC code repositories for
use in fine-tuning LLMs or including in prompt context for few-shot learning.}

%% file: ack.tex
This material is based upon work supported in part by the National Science
Foundation (NSF) under Grant No.~2047120, and the NSF Graduate Research
Fellowship Program under Grant No.~DGE~2236417. This material is based upon
work supported in part by the U.S.~Department of Energy (DOE), Office of
Science, Office of Advanced Scientific Computing Research, through solicitation
DE-FOA-0003264, ``Advances in Artificial Intelligence for Science'', under
Award Number DE-SC0025598. This work was performed under the auspices of the
U.S.~DOE by Lawrence Livermore National Laboratory under
Contract~DE-AC52-07NA27344 (LLNL-CONF-855581).

This research is supported by the National Artificial Intelligence Research
Resource (NAIRR) Pilot and the Delta advanced computing and data resource which
is supported by the NSF (award NSF-OAC 2005572) and the State of Illinois.  The
authors acknowledge the University of Maryland supercomputing resources made
available for conducting the research reported in this paper.